\documentclass{iopart}
\include{epsf}
\begin{document}

\title{Frustration and Melting of 
Colloidal Molecular Crystals}
\author{C J Olson Reichhardt\dag\ and C Reichhardt\ddag}

\address{\dag T-12 and \ddag CNLS, Theoretical Division, Los Alamos National
Laboratory, Los Alamos, NM 87545, USA}

\begin{abstract}
Using numerical simulations we show that a  variety of
novel colloidal crystalline states
and multi-step melting phenomena occur on square and
triangular two-dimensional periodic substrates. At 
half-integer fillings different kinds of frustration effects can
be realized.  
A two-step melting transition
can occur in which individual colloidal molecules initially rotate,
destroying the overall orientational order, followed by the onset
of interwell colloidal hopping, in good agreement with recent 
experiments.  
\end{abstract}

\submitto{\JPA}


\nosections
Colloidal particles are an ideal system for studying 2D ordering and
melting for different kinds of substrates as the 
individual particle positions and dynamics can
be directly visualized. 
Crystallization and melting on 2D periodic substrates is 
also relevant to vortices in superconductors with periodic 
pinning arrays \cite{Baert,Harada,Reichhardt} and
atomic orderings \cite{Coppersmith}. 
Colloidal crystallization on 2D periodic
substrates has been the subject of considerable recent interest.
In several recent experimental studies, a 2D substrate for colloids
was created using optical tweezer arrays \cite{Dufresne}, templating
\cite{Lin,Pertsinidis}, and 2D crossed laser arrays \cite{Bechinger}.
Colloidal ordering and melting on 2D periodic square and triangular
substrates has been studied through simulation \cite{PRL} and
experiment \cite{BechingerPRL}.  It was shown that a rich variety of novel
colloidal crystalline states, referred to as colloidal molecular
crystals (CMC's), form at integer filling of the periodic substrates.

In Ref. \cite{PRL}, integer matching states up through a filling of four
colloids per minima were illustrated for both square and triangular
arrays, and the two-stage melting behavior of the dimer state at
a filling of two colloids per minima was demonstrated for the square
substrate.  Here, we present the structures observed at half-integer
filling for the square and triangular substrates, showing that the perfect
regularity of the integer filling state is destroyed away from integer
filling.  We also demonstrate a two-stage melting transition for
the trimer state at a filling of three colloids per minima for the
square substrate.

We simulate a 2D system of $N_c$ colloids with periodic boundary conditions
in the $x$ and $y$ directions, using Langevin dynamics as employed in
previous colloidal simulations \cite{Schweigert}.  The overdamped 
equation of motion for a colloid $i$ is

\[ \frac{d{\bf r}_{i}}{dt}={\bf f}_{i} + {\bf f}_{s} + {\bf f}_{T}. \]

Here ${\bf f}_{i} = -\sum_{i\ne j}^{N_{c}} \nabla_{i} V(r_{ij})$ is
the interaction force from the other colloids.  The colloid-colloid 
interaction is a Yukawa or screened Coulomb potential, 
$V(r_{ij})=(Q^{2}/|{\bf r}_{i}-{\bf r}_{j}|)\exp(-\kappa |{\bf r}_{i}
- {\bf r}_{j}| )$, where $Q=1$ is the charge of the particles,
$1/\kappa$ is the screening length, and ${\bf r}_{i(j)}$ is the
position of particle $i$ ($j$).  The system length is measured in units
of the lattice constant $a_{0}$ and we take the screening length
$1/\kappa = a_{0}/2$.  For the force from the 2D substrate, we consider
both square and triangular substrates with strength A, period $a_{0}$,
and $N_{m}$ minima.  For square substrates, 
${\bf f}_{s}= A\sin(2\pi x/a_{0})\hat{x} + A\sin(2\pi y/a_{0})\hat{y}$,
and for triangular substrates,
${\bf f}_{s}=\sum_{i=1}{3} A \sin(2 \pi p_i / a_0) [\cos(\theta_i){\bf \hat{x}}
- \sin(\theta_i){\bf \hat{y}}]$,
where
$p_i = x \cos(\theta_i) - y \sin(\theta_i) + a_0 / 2$,
$\theta_1 = \pi/6$, $\theta_2 = \pi/2$, and 
$\theta_3 = 5\pi/6$.
The thermal force ${\bf f}_{T}$ is a randomly fluctuating force
from random kicks.  We start the system at a temperature where all
the colloids are diffusing rapidly and gradually cool to $T=0.0$.  
We do not take into account hydrodynamic effects or possible long-range
attractions between colloids.

The colloidal positions for a system with a square substrate at
integer matching between the colloidal periodicity and the substrate
periodicity, $N_c = n N_m$, are illustrated in Fig. 1 of Ref. \cite{PRL}
for $n = 1$ to 4.  In particular, at $N_c = N_m$, each colloid is
located at the center of the potential minima and a square colloidal
crystal forms, while at $N_c = 2N_m$, each minima captures two colloids
which can be regarded as a colloidal dimer with a rotational degree
of freedom.  Over a range of substrate strengths, the colloidal dimers
form a rotationally ordered state, with neighboring dimers
perpendicular to one another.  The orientational ordering of the
dimers is due to the colloidal repulsion, and allows the distance between
the colloids to be maximized under the constraint of the square 
substrate.
In Fig. 1, we show the colloidal positions for
the square 
substrate system at a half-integer filling, $N_c = 1.5 N_m$.  Here
we find a combination of the single colloids and dimer colloids. 
The single colloids are located at every other site with intervening
dimers, creating a checkerboard ordering.
The strong orientational ordering of the dimers observed at $N_c = 2N_m$
is lost in this state.  The dimers alternate their orientation between
vertical and horizontal in a disordered fashion, forming a pattern of
grain boundaries.

The ordered colloidal crystalline states that form on a triangular
substrate at integer matching from $n=1$ to 4 are illustrated
in Fig. 2 of Ref. \cite{PRL}.  As in the case of the square substrate,
at $N_c=N_m$ each minima captures a single colloid, while at
$N_c=2N_m$ each minima captures two colloids which form a dimer
state.  The dimers again have an additional orientational ordering
in which the dimers in each row have the same orientation, which is
rotated $45^\circ$ with respect to the adjacent rows.  
In Fig. 2 we illustrate the colloidal positions for the triangular
substrate at half-integer filling, $N_c=1.5N_m$.  Unlike the square
substrate, where dimers could tile the lattice in an orderly way,
filling every other minima, for the triangular substrate the 
positioning of the dimers is frustrated.  It is not possible
to arrange a state such that minima with only one colloid will
be between every other dimer.
As a result, the 
colloids form a very disordered state in which the dimers show a wide
range of orientations, and not merely two as in the case of the square
substrate at $N_c=1.5N_m$, or the triangular substrate at 
$N_c=2N_m$.

In Fig. 3 we show the two-stage melting of the CMC at $N_c=3N_m$ on
the square lattice.  As illustrated in Fig. 3(a) at low
temperatures $T/T_{m}^{0} < 0.25$ (where $T_{m}^{0}$ is the
melting temperature at zero substrate strength), both orientational
and translational order of the trimers are present and the system is
frozen.  This is the ``ordered solid'' phase.  In Fig. 3(b), 
at $T/T_{m}^{0}=1.5$, the trimers begin to rotate within the minima;
however, diffusion of individual colloids throughout the sample does
not occur.  The system is still frozen but the trimer orientational
order is lost.  This is the ``partially ordered solid'' phase.
In Fig. 3(c), for a higher temperature $T/T_{m}^{0}$, the system
enters a modulated liquid phase.  Here the colloids begin to 
diffuse throughout the system. We note that in the 
recent experiments of Brunner and Bechinger \cite{BechingerPRL},
the same melting phenomena was also found for trimers on a
triangular substrate.

To summarize, we have shown the rich variety of novel colloidal
crystalline and disordered states that can be achieved with square
and triangular two-dimensional substrates.  The colloidal molecular
crystal states that appear at integer filling of the substrate
minima persist at half-integer filling, appearing as a mixture of
the two neighboring integer filling states.  For the case of
$N_c=1.5N_m$, the single and dimer states coexist.  An ordered
filling of dimers can be arranged on the square substrate at
$N_c=1.5N_m$, but the dimer arrangement is frustrated on the
triangular substrate and a strongly disordered state results.
The triangular lattice at $N_c=1.5N_m$ is thus a
realization of a frustrated system which would provide an interesting
topic for further study.  For integer filling of $N_c=3N_m$ on
a square substrate, we demonstrate
the multistage melting of the CMC, where the orientational order of
the colloidal molecule states is lost first, followed by the
translational order.  Since the colloids within a minima can act
as a single particle with a rotational degree of freedom, our
results also suggest that certain canonical statistical mechanics
models, such as Ising, $XY$, Potts, and frustrated models, may
be realized with colloids on two-dimensional periodic substrates.
The states predicted here should be observable for colloids
interacting with crossed-laser arrays or optical tweezer arrays,
dusty plasmas in 2D with periodic potentials, and vortices in
superconductors with periodic substrates.

\ack{
We thank C. Bechinger, M. Brunner, 
D.G. Grier, P. Korda, X.S. Ling, A. Pertsinidis,
and G. Spalding for useful discussions.
This work was supported by the U.S. DoE under
Contract No. W-7405-ENG-36.}


\begin{figure}
\begin{center}
\epsfxsize=3.5in
\epsfbox{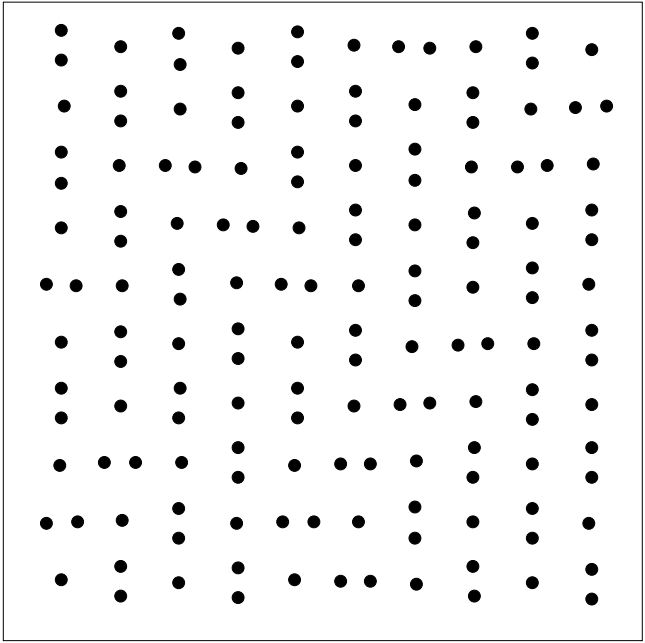}
\end{center}
\caption{The colloid configurations (black dots) at $T=0.0$ for a square
2D periodic substrate with $A=2.5$, for a half-integer colloidal
density, $N_c=1.5 N_m$.  Every other minima captures a single colloid,
while the remaining minima capture two colloids in a dimer state.  There is
no long-range rotational order of the dimers.}
\end{figure}

\begin{figure}
\begin{center}
\epsfxsize=3.5in
\epsfbox{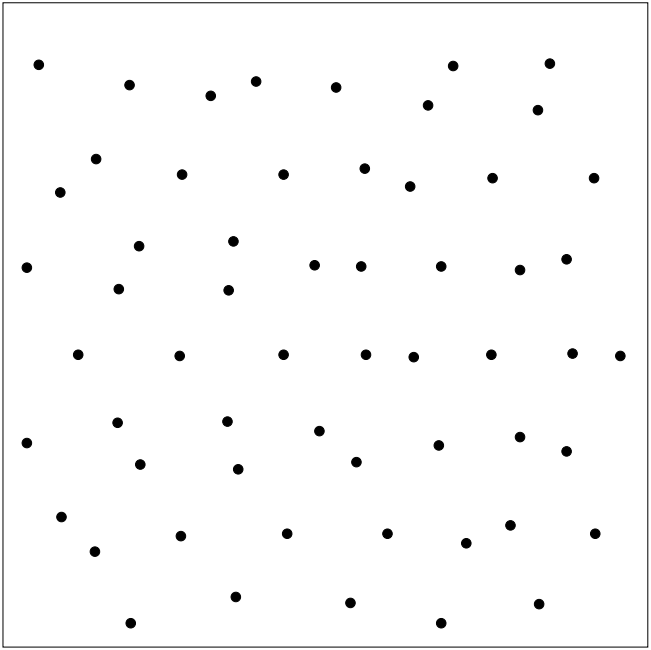}
\end{center}
\caption{The colloid configurations (black dots) at $T=0.0$ for a 
triangular
2D periodic substrate with $A=2.5$, for a half-integer colloidal
density, $N_c=1.5 N_m$.  It is not possible to arrange
a state such that minima with only one colloid will be between
every other dimer.  Therefore there is no dimer ordering.}
\end{figure}

\begin{figure}
\begin{center}
\epsfxsize=3.5in
\epsfbox{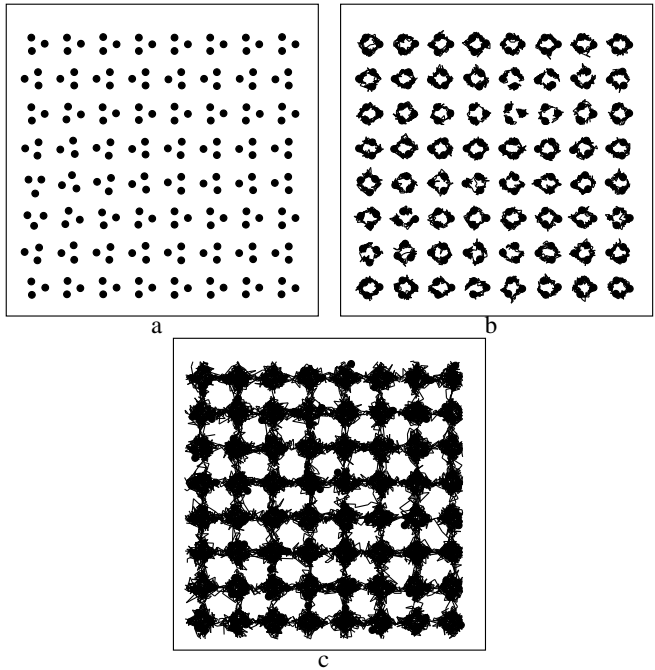}
\end{center}
\caption{The colloid positions (black dots) and trajectories (lines) over
fixed time intervals at different temperatures for the trimer state
on a square substrate shown in Fig. 1(c) of Ref. \cite{PRL}.  (a) The
ordered solid phase at $T/T_{m}^{0}=0.25$, where $T_{m}^{0}$ is the
melting temperature at zero substrate strength.  (b) $T/T_{m}^{0}=1.5$.
Orientational order is destroyed as the trimers rotate within the
substrate minima, but the colloids remain trapped inside each minima
so the system is still in a solid phase.  (c) $T/T_{m}^{0}=4.0.$  
Individual colloidal diffusion occurs throughout the sample and the
system is in the liquid phase.}
\end{figure}

\end{document}